# Guessing probability distributions from small samples

Thorsten Pöschel [a,b,1] Werner Ebeling [a,2] Helge Rosé [a,3]

[a] *Humboldt–Universität zu Berlin, Institut für Physik, Unter den Linden 6, D–10099 Berlin, Germany, http://summa.physik.hu-berlin.de:80/~thorsten/*

[b] *The James Franck Institute, The University of Chicago, 5640 South Ellis Av., Chicago, Illinois 60637*

**Abstract**

We propose a new method for the calculation of the statistical properties, as e.g. the entropy, of unknown generators of symbolic sequences. The probability distribution $p(k)$ of the elements $k$ of a population can be approximated by the frequencies $f(k)$ of a sample provided the sample is long enough so that each element $k$ occurs many times. Our method yields an approximation if this precondition does not hold. For a given $f(k)$ we recalculate the Zipf–ordered probability distribution by optimization of the parameters of a guessed distribution. We demonstrate that our method yields reliable results.

## 1 Introduction

Given a statistical population of discrete events $k$ generated by a stationary dynamic process, one of the most interesting statistical properties of the population and hence of the process is its entropy. If the sample space, i.e. the number of different elements which are allowed to occur in the population, is small compared with the size of a drawn sample one can approximate the probabilities $p(k)$ of the elements $k$ by their relative frequencies $f(k)$ and one finds for the observed entropy $H_{obs}$

$$H = -\sum_k p(k) \log p(k) \approx -\sum_k f(k) \log f(k) = H_{obs} \ . \qquad (1)$$

[1] E–mail: thorsten@hlrsun.hlrz.kfa–juelich.de
[2] E–mail: werner@itp02.physik.hu–berlin.de
[3] E–mail: rose@summa.physik.hu–berlin.de

If the number of the allowed different events is not small compared with the size of the sample the approximation $p(k) \approx f(k)$ yields dramatically wrong results. In this case the knowledge of the frequencies is not sufficient to determine the entropy. The aim of this paper is to provide a method to calculate the entropy and other statistical characteristics for the case that the approximation (1) does not hold.

An interesting example of such systems are subsequences (words) of length $n$ of symbolic sequences of length $L$ written using an alphabet of $\lambda$ letters. Examples are biosequences like DNA ($\lambda = 4$, $L \lesssim 10^9$), literary texts ($\lambda \approx 80$ letters and punctuation marks, $L \lesssim 10^7$) and computer files ($\lambda = 2$, $L$ arbitrary). For the case of biosequences there is a variety of $\lambda^n = 1,048,576$ different words of length $n = 10$. To measure the probability distribution of the words directly by counting their frequencies we need at least a sequence of length $10^8$ to have reliable statistics. Therefore the ensemble of subsequences of length $n$ is a typical example where the precondition does not hold. To illustrate the problem we calculate the observed entropy $H_{obs}^{(n)}$ for $N$ words of length $n$ in a Bernoulli sequence with $\lambda = 2$ where both symbols occur with the same probability. The exact result is $H^{(n)} = n \log \lambda$. In figure 1 we have drawn the values of $H^{(n)}$ and $H_{obs}^{(n)}$ over $n$. The observed entropy values are correct for small word length $n$ when we can approximate the probabilities by the relative frequencies. For larger word length, however, the observed entropies are significantly below the exact values, even for very large samples (circles: $N = 10^6$, diamonds: $N = 10^4$).

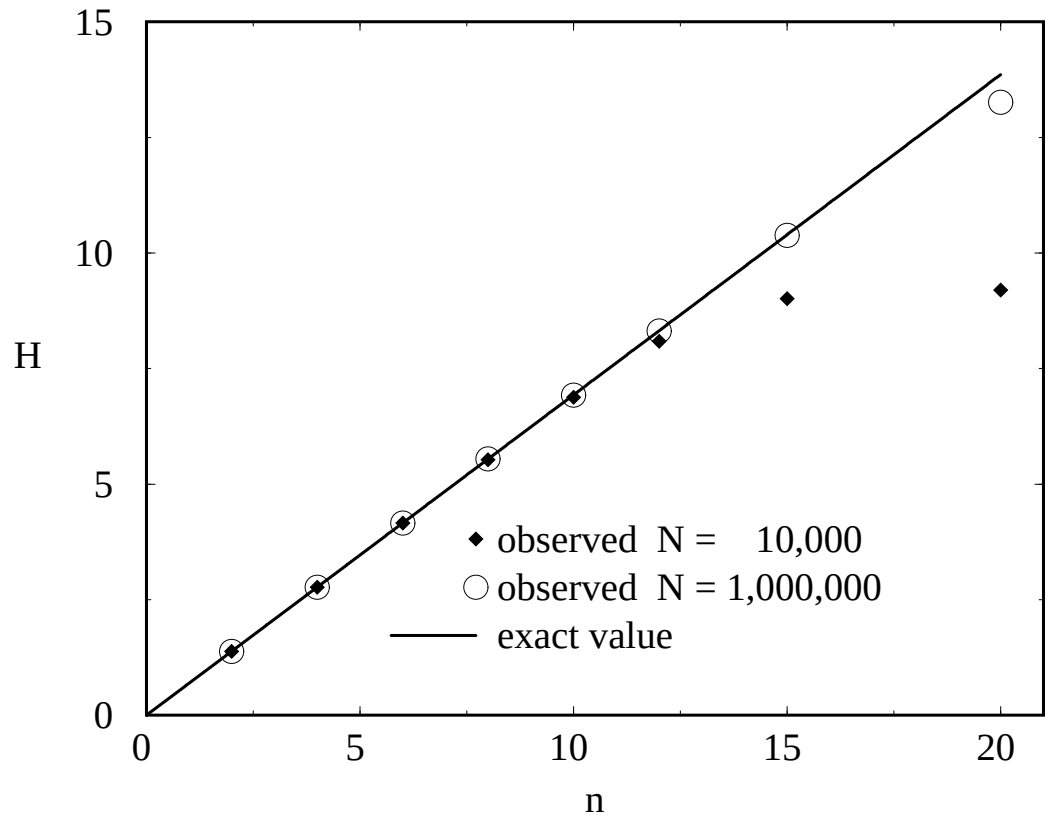

Fig. 1. The observed entropy for $N$ words of length $n$ from a Bernoulli sequence.

Under several strong preconditions the probabilities of words in sequences can be estimated from the frequencies using various correction methods [1–3]. The advanced algorithm proposed in [3] is based on a theorem by McMillan and Khinchin [4] saying that for word length $n \to \infty$ the frequencies of the admitted substrings of a sequence are equally distributed. If one is interested in the entropies for *finite* words, however, the theoretical basis to apply this theorem is weak and there is no evidence about the reliability of the results. More-over this theorem is proven for Markov sequences only. In sequences gathered from natural languages, biosequences and other natural or artificial sources it is very unlikely that the probabilities of the words of interesting length, e.g. words or sentences for languages, amino acids or elements of the hidden "DNA language" for biosequences, are equally distributed. Otherwise we had to assume that all English five–letter–words are equally frequent. Certainly this is not the case.

## 2 Description of the method

To calculate the entropy of a distribution it is not necessary to determine for each event $k$ the probability $p(k)$. It is sufficient to determine the values of the probabilities without knowing which probability belongs to which event. Generally spoken if we assume to have $K$ events there are $K!$ different relations $k \leftrightarrow p$. We need not to determine one particular (the correct relation) but only one arbitrary of them. Hence the calculation of the entropy is $K!$ times easier than to determine the probability $p(k)$ for each event $k$. We assume a special order where the first element has the largest probability, the second one the second largest etc. We call this distribution Zipf–ordered. Zipf ordering means that the probabilities of the elements are ordered according to their rank and therefore the distribution $p(k)$ is a monotonically decaying function. The following procedure describes a method how to reconstruct the Zipf–ordered probability distribution $p(k)$ from a finite sample.

Provided we have some reason to expect (to guess) the parametric form of the probability distribution. As an example we use a simple distribution $p(k, \alpha, \beta, \gamma)$ with $k = 1, 2, \ldots$ consisting of a linearly decreasing and a constant part

$$p(k) = \begin{cases} \frac{2-\alpha\beta}{2\gamma} + \alpha\,(1 - \frac{k}{\beta}) : 1 \leq k < \beta \\ (2 - \alpha\,\beta)/(2\,\gamma) : \beta \leq k \leq \gamma \\ 0 : k > \gamma \quad . \end{cases} \tag{2}$$

Then the algorithm runs as follows:

i. Find the frequencies $F(k)$ for the $N$ events $k$ and order them according to their value (Zipf–order). The index $k$ runs over all *different* events occurring in the sample ($k \in \{1 \ldots K^{MAX}\}$). Note: there are $N$ events but only $K^{MAX}$ different ones. Normalize this distribution $F_1(k) = F(k)/N$. There are various sophisticated algorithms to find the frequencies of large samples and to order them (e.g. [5]). As in earlier papers [6] we applied for finding the elements a "hashing"–method and for sorting a mixed algorithm consisting of "Quicksort" for the frequent elements and "Distribution Counting" for the long tail of elements with low frequencies.

ii. Guess initial conditions for the parameters (in our case $\alpha$, $\beta$ and $\gamma$).

iii. Generate $M$ samples of $N$ random integers ($RI_k^m$, $k = 1 \ldots N$, $m = 1 \ldots M$) according to the parametric probability distribution $p(k, \alpha, \beta, \gamma)$. In the following examples we used $M = 20$. Order each of the samples according to the ranks $f_i(k, \alpha, \beta, \gamma)$ $(i = 1 \ldots M)$. Average over the $M$ ordered samples

$$\overline{f(k, \alpha, \beta, \gamma)} = \frac{1}{M} \sum_{i=1}^{M} f_i(k, \alpha, \beta, \gamma) \tag{3}$$

with $k \in \{1, k^{max}\}$ and $k^{max} = \max\left(k_i^{max},\ \ (i = 1 \ldots M)\right)$. Since we want to determine the averaged or typical Zipf–ordered distribution, it is important to order the elements first and then to average. Normalize the averaged distribution of the frequencies

$$\overline{f_1(k, \alpha, \beta, \gamma)} = \left( \sum_{k=0}^{k^{max}} \overline{f(k, \alpha, \beta, \gamma)} \right)^{-1} \overline{f(k, \alpha, \beta, \gamma)}\ . \tag{4}$$

iv. Measure the deviation $D$ between the normalized averaged simulated frequency distribution $\overline{f_1(k, \alpha, \beta, \gamma)}$ and the frequency distribution $F_1(k)$ of the given sample according to a certain rule, e.g.

$$D = \sum_{k=1}^{K} \left( \frac{\overline{f_1(k, \alpha, \beta, \gamma)}}{F_1(k)} - 1 \right)^2 \ ,\ \ K = \max\left\{ k^{max}, K^{MAX} \right\}\ . \tag{5}$$

v. Change the parameters of the guessed probability distribution $p(k)$ (in our case the parameters $\alpha$, $\beta$ and $\gamma$) due to an optimization rule (e.g. [7]) which minimizes $D$ and proceed with the third step until the deviation $D$ is sufficiently small.

vi. Extract the interesting statistical properties out of the probability distribution $p(k)$ using the parameters $\alpha^*$, $\beta^*$ and $\gamma^*$ which have been gathered during the optimization process.

## 3 Examples

### *3.1 Entropy of artificial sequences*

We generated a statistical ensemble $N = 10^4$ according to the probability distribution eq. (2) with $\alpha = 9.0 \cdot 10^{-6}$, $\beta = 10,000$ and $\gamma = 50,000$. Fig. 2 (solid lines) shows the probability distribution $p(k)$ and the Zipf–ordered frequencies $f(k)$.

Optimizing the parametric guessed probability distribution using the proposed method we find for the optimized parameters $\alpha^* = 9.22 \cdot 10^{-6}$, $\beta^* = 12,900$ and $\gamma^* = 50,000$, i.e. the guessed and the actual distributions fall almost together. Since we know the original probability distribution (eq. 2) we can compare its exact entropy with the entropy of the guessed probability distribution $H_{guess}$ and with the observed entropy $H_{obs}$ due to eqs. (6,7).

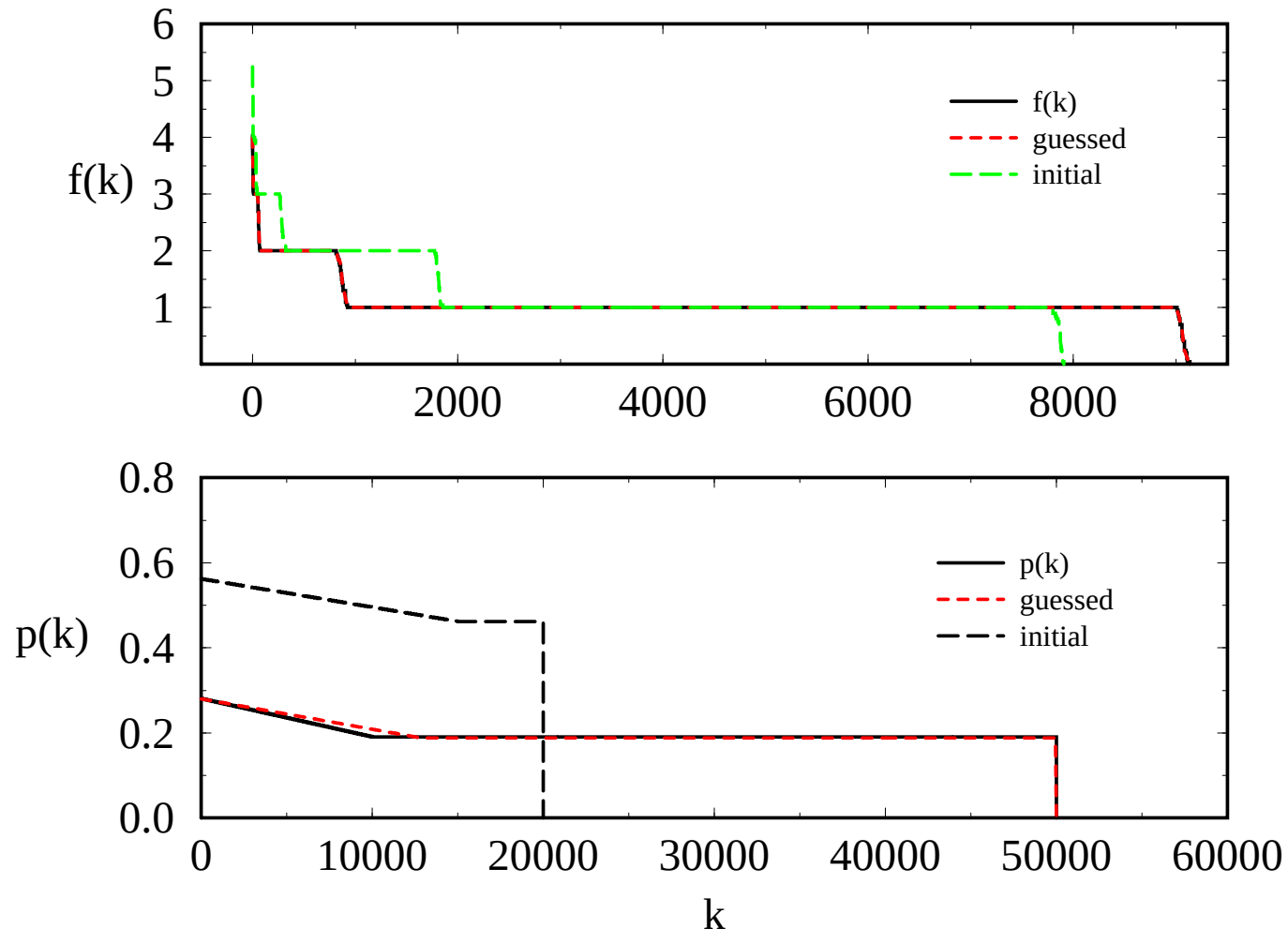

Fig. 2. The probability distribution $p(k)$ (eq. 2) and the Zipf–ordered frequencies $f(k)$ corresponding to this distribution. The dashed lines which can almost not be distinguished from the solid lines display the guessed distributions. The initial distributions before optimization is drawn with wide dashes.

$$H_{guess}(k, \alpha^*, \beta^*, \gamma^*) = -\sum p(k, \alpha^*, \beta^*, \gamma^*) \cdot \log p(k, \alpha^*, \beta^*, \gamma^*) \tag{6}$$

$$H_{obs}(k) = -\sum F_1(k) \cdot \log F_1(k) \tag{7}$$

We found $H_{obs} = 9.0811$ and $H_{guess} = 10.8147$, the exact value according to $p(k, \alpha, \beta, \gamma)$ (eq. 2) is $H = 10.8188$.

Now we try to guess a probability of a more complicated form

$$p(k) = \begin{cases} \alpha\,(k - \epsilon)^{-\frac{1}{3}} : 1 \le k < \beta \\ \phi\, k^{-\delta} : \beta \le k \le \gamma \\ 0 : k > \gamma\ . \end{cases} \tag{8}$$

(As we will show below this function approximates the probability distribution of the words in an English text.) The variables $\alpha$ and $\phi$ can be eliminated due to normalization and continuity condition. The test sample of size $N = 10^4$ was generated using $\epsilon = 0.9$, $\beta = 22$, $\delta = 0.64$ and $\gamma = 70,000$. After the optimization we guess the parameters $\epsilon^* = 0.79$, $\beta^* = 21.9$, $\delta^* = 0.63$ and $\gamma^* = 65,000$. Fig. 3 shows the original and the guessed probability distributions and the Zipf–ordered frequencies for both cases. The guessed entropy $H_{guess} = 10.5053$ approximates the exact value $H = 10.5397$ very well while the observed entropy $H_{obs} = 8.8554$ shows a clear deviation from the correct value.

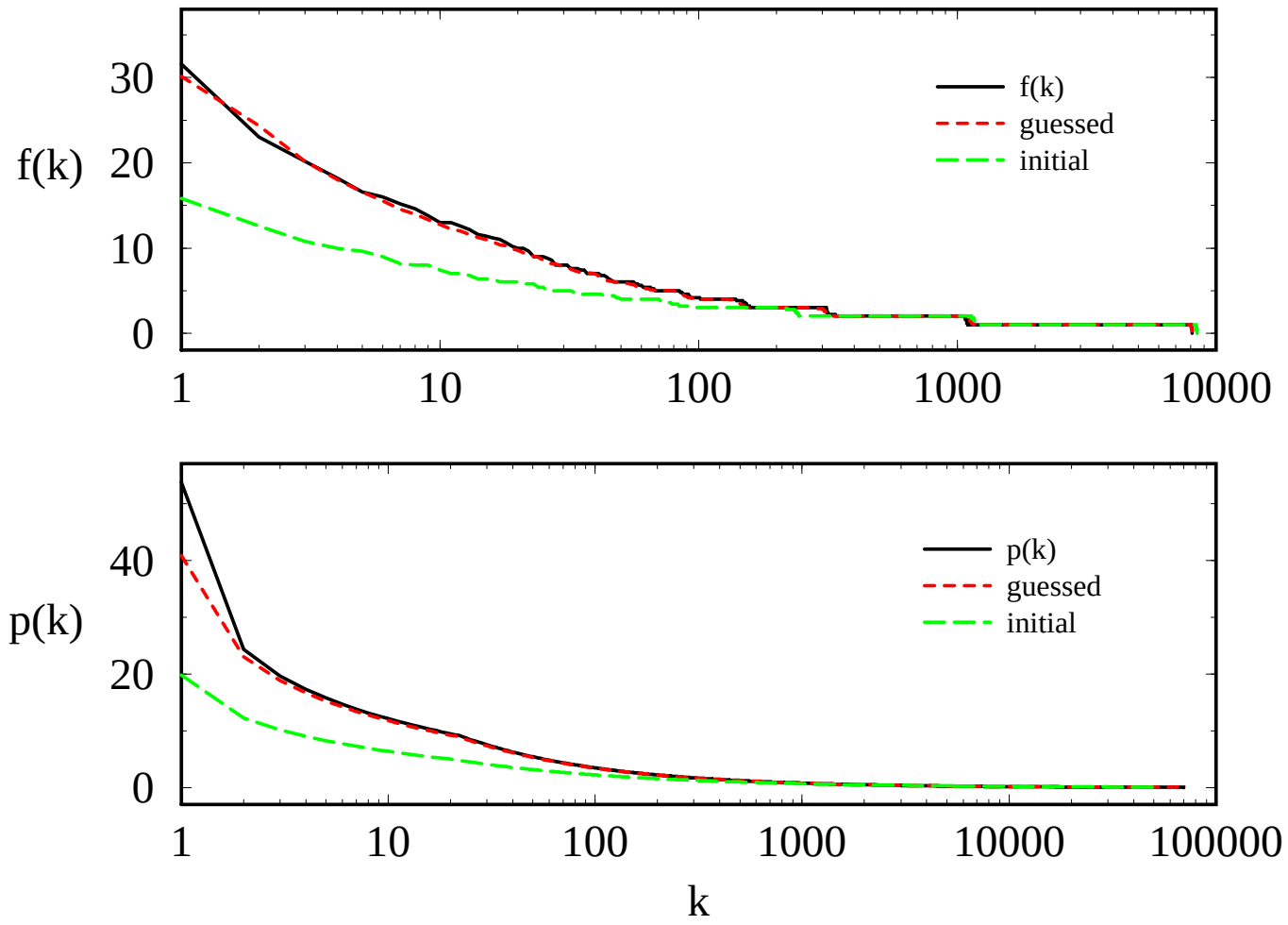

Fig. 3. The original and guessed probability distributions and the Zipf–ordered frequencies for the distribution in eq. (8) ($N = 10^4$).

### *3.2 Words in an English text*

With the ansatz (8) we tried to guess the probability distribution of the words of different length $n$ in the text “Moby Dick” by H. Melville [8]. The text was mapped to an alphabet of $\lambda = 32$ letters as described in [9]. Depending on overlapping or non–overlapping counting of the words we expect different results. We note that overlapping counting is statistically not correct since the elements of the sample are not statistically independent, however, only overlapping counting yields enough words to get somehow reasonable results for the observed entropy. We will show that our method works in both cases, overlapping and non–overlapping. Fig. 4 shows the ordered frequencies of $N = 5 \cdot 10^4$ words of the length $n = 6$. The optimized distribution eq. (8) reproduces the original frequency distribution (Moby Dick) with satisfying accuracy.

Using the ansatz (8) we found $\epsilon^* = 0.73$, $\beta^* = 31$, $\delta^* = 0.70$ and $\gamma^* = 129,890$. This calculation was carried out for various word lengths $n$. Fig. 5 shows the entropies $H^{(n)}_{obs}$ and $H^{(n)}_{guess}$ according to eqs. (6,7) as a function of $n$. All results obtained have been derived from a set of $5 \cdot 10^4$ non-overlapping words taken from the text of length $L = 10^6$. When we count overlapping words, we find surprisingly that the entropy is quite insensitive (see curves using filled and empty diamonds in fig. 5). The rather difficult problem of overlapping or non–overlapping counting will be addressed in detail in [10]. Since the exact probability distribution for the words in "Moby Dick" is unknown we compare the guessed entropy (crosses) with the observed entropy (empty diamonds: overlapping counting, full diamonds: non-overlapping counting) and an estimation of the entropy using an extrapolation method (see [6]), all based on the same set of data ($N = 5 \cdot 10^4$), and with the observed entropy based on a twenty times larger set of data (boxes: overlapping counting). As expected, for longer word length $n$ the observed values $H^{(n)}_{obs}$ underestimate the entropy. For small $n$ they are reliable due to the reliable statistics. The guessed entropy $H^{(n)}_{guess}$ agrees for small $n$ with the observed entropy and for large $n$ with the extrapolated values.

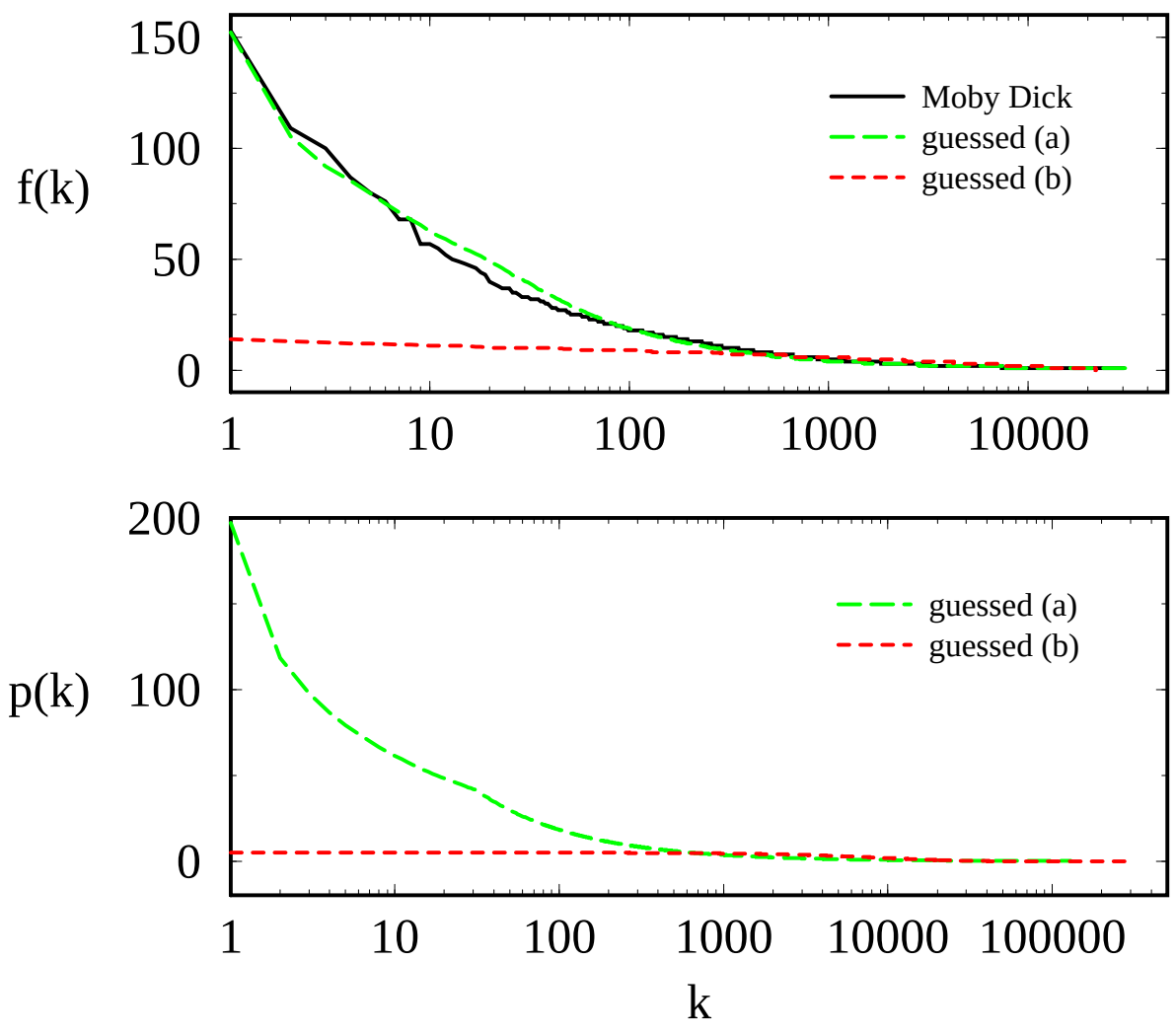

Fig. 4. Zipf–ordered frequencies of words of length $n = 6$ in "Moby Dick". The curves guessed (a) and (b) (top) display the frequency distributions which have been reproduced using the guessed probability distributions in the bottom figure according to eq. (8) (a) and eq. (9) (b).

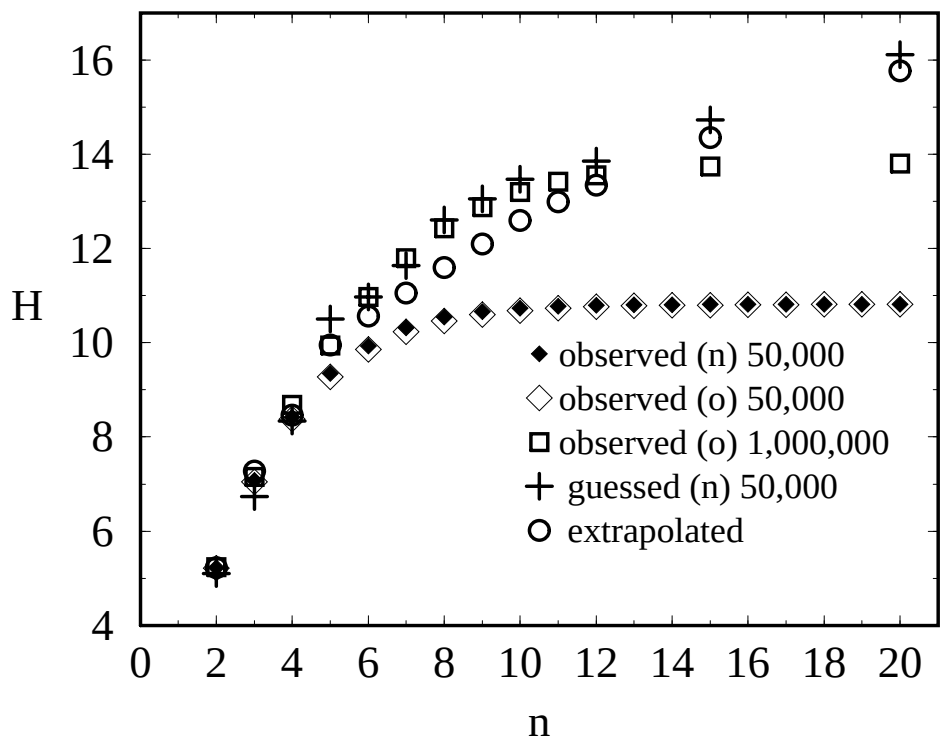

Fig. 5. *Observed entropy* $H_{obs}^{(n)}$ *and guessed entropy* $H_{guess}^{(n)}$ *for the text "Moby Dick"* *(*$N = 5 \cdot 10^4$*) over the word length* $n$*. The circles* $\circ$ *display the results of an extrapolation method described in [6] and the boxes* $\square$ *show the observed entropy of the text using* $N = 10^6$ *words. (o) denotes overlapping and (n) non-overlapping counting.*

The form of the guessed theoretical distribution $p(k, \alpha, \beta, \gamma, \ldots)$ is arbitrary as long as it is a normalized monotonically decreasing function (Zipf-order). Suppose that one has no information about the mechanism which generated a given sample. Then one has to find the functional form of the guessed distribution which is most appropriate to a given problem, i.e. in the ideal case the guessed distribution contains the real probability distribution as special case without being too complicated. An ansatz $p(k, \alpha, \beta, \gamma, \ldots)$ is suited if the optimized guessed probability distribution reproduces the frequency distribution of the original sample with satisfactory accuracy.

The ansatz (8) looks rather artificial: in fact we tried several forms of the guessed probability distribution and the one proposed in eq. (8) turned out to be the best of them. None of the others reproduces the frequencies sufficiently correct. For demonstration we assume the function

$$p(k, \alpha, \beta) = \begin{cases} \alpha \cdot \exp(-\beta\, k) : k \leq \gamma \\ 0 : k > \gamma \end{cases} \tag{9}$$

with the normalization $\gamma = -\beta^{-1} \log(1 - \beta/\alpha)$. The optimized function is drawn in figure 4 (*guessed (b)*). We find that the frequency distribution reproduced from this function differs much more from the original frequency distribution (Moby Dick) than that of the guess according to eq. (8).

Admittedly any similar ansatz showing a well pronounced peak for low ranks *(frequent words)*, a long plateau with slow decrease *(standard words)* and a long tail *(seldom words)*, could give reliable results as well. Anyhow there is no wide choice for the parametric form of the probability distribution. Eq. (8) belongs to the class of distributions fulfilling this *three–region criterion.* For a more detailed discussion of the statistics of words see e.g. [11] and many references therein.

## 4 Discussion

The problem addressed in this paper was to find the rank ordered probability distribution from the given frequency distribution of a finite sample. For finite samples (Bernoulli sequence and English text) we have shown that the calculation of the entropy using the relative frequencies instead of the (unknown) probabilities yields wrong results.

We could show that the proposed algorithm is able to find the correct parameters of a guessed probability distribution which reproduces the statistical characteristics of a given symbolic sequence. The method has been tested for samples generated by well defined sources, i.e. by known probability distributions, and for an unknown source, i.e. the word distribution of an English text. For the sample sequences we have evidence that the algorithm yields reliable results. The deviations of the entropy values from the correct values are rather small and in all cases far better than the observed entropies. For the unknown source "Moby Dick" we have no direct possibility to check the quality of the method, however, the calculated entropy values agree for small word lengths $n$ with the observed entropy and for larger $n$ with the results of an extrapolation method [6]. In this sense both approaches support each other. The proposed algorithm can be applied to the trajectories of dynamic systems. Formally the trajectory of a discrete dynamics is a text written in a certain language using $\lambda$ different letters. The rank ordered distribution of sub-trajectories of length $n$ belongs to the most important characteristics of a discrete dynamic system. In this way we consider the analysis of English text as an example for the analysis of a very complex dynamic system.

In many cases there is a principal limitation of the available data, i.e. the available samples are small with respect to the needs of a reliable statistics, and hence there is a principal limitation for the calculation of the statistical properties using frequencies instead of probabilities. For such cases using the proposed method one can calculate values which could not be found so far. Given a finite set of data the proposed method yields the most reliable values for the Zipf–ordered distributions and the entropies which are presently available. The method is not restricted to the calculation of the entropy but all statistical properties which depend on the Zipf–ordered probability distribution can be estimated using the proposed algorithm.

## Acknowledgement

We thank T. Aßelmeyer, H. J. Herrmann and L. Schimansky–Geier for discussion and the *Project Gutenberg Etext, Illinois Benedictine College, Lisle* for providing the ASCII–text of "Moby Dick".